\documentclass[a4paper,11pt]{article}
\pdfoutput=1 
\usepackage{jheppub}
\usepackage[T1]{fontenc} 
\usepackage{color}
\usepackage{fancybox}
\usepackage{amsmath}
\usepackage{amsfonts}
\usepackage{slashed}
\usepackage{amssymb}
\usepackage{indentfirst}
\usepackage{stmaryrd}
\usepackage{subcaption}
\usepackage{graphicx}
\usepackage{cleveref}
\usepackage{textcomp}
\usepackage[toc]{appendix}
\usepackage[font={small,it}]{caption}
\usepackage[justification=centering]{caption}

\allowdisplaybreaks

\newcommand{\be}{\begin{equation}}
\newcommand{\bea}{\begin{eqnarray}}

\newcommand{\ee}{\end{equation}}
\newcommand{\eea}{\end{eqnarray}}

\title{\boldmath Equations of motion from Cederwall's pure spinor superspace actions}

\author{Nathan Berkovits$^{*}$, Max Guillen$^{*}$}

\affiliation{$^{*}$ICTP South American Institute for Fundamental Reserch\\
Instituto de F\'{i}sica Te\'{o}rica, UNESP-Universidade Estadual Paulista\\ R. Dr. Bento T. Ferraz 271, Bl. II, S\~{a}o Paulo 01140-070, SP, Brazil}

\emailAdd{nberkovi@ift.unesp.br, luismax@ift.unesp.br}

\abstract{Using non-minimal pure spinor superspace, Cederwall has constructed BRST-invariant actions for $D=10$ super-Born-Infeld and $D=11$ supergravity which are quartic in the superfields. But since the superfields have explicit dependence on the non-minimal pure spinor variables, it is non-trivial to show these actions correctly describe super-Born-Infeld and supergravity. In this paper, we expand solutions to the equations of motion from Cederwall's actions to leading order around the linearized solutions and show that they correctly describe the interactions of $D=10$ super-Born-Infeld and $D=11$ supergravity. 
}


\begin{document} 
\hfill{}
\maketitle

\section{Introduction}

Pure spinors $\lambda^\alpha$ in ten and eleven dimensions have been useful for constructing vertex operators and computing on-shell scattering amplitudes with manifest spacetime supersymmetry in super-Yang-Mills, supergravity and superstring theory \cite{HOWE1991141,Berkovits:2001rb,Berkovits:2002uc,Gomez:2013sla}.
After including non-minimal variables $(\bar\lambda_\alpha, r_\alpha)$, pure spinors have also been useful for constructing BRST-invariant off-shell actions for these maximally supersymmetric theories \cite{Cederwall:2010tn,Cederwall:2013vba,Cederwall:2011vy}.

These BRST-invariant actions have a very simple form and were constructed by Cederwall using superfields $\Psi(x^m, \theta^\alpha, \lambda^\alpha, \bar\lambda_\alpha, r_\alpha)$ which transform covariantly under spacetime supersymmetry and depend on both the usual superspace variables $(x^m, \theta^\alpha)$ and the non-minimal pure spinor variables $(\lambda^\alpha, \bar\lambda_\alpha, r_\alpha)$. Although the actions require a non-supersymmetric regulator to define integration over the non-minimal pure spinor variables, it is easy to show that the supersymmetry transformation of the regulator is BRST-trivial so the action is spacetime supersymmetric.

However, since the superfields $\Psi$ can depend in a non-trivial manner on the non-minimal variables, it is not obvious how to show that the solutions to the equations of motion correctly describe the usual on-shell $D=10$ and $D=11$ superfields which depend only on the $(x^m, \theta^\alpha)$ superspace variables.

In this paper, an explicit procedure will be given for extracting the usual on-shell
$D=10$ and $D=11$ superfields from the equations of motion of the pure spinor actions for the cases of $D=10$ supersymmetric Born-Infeld and for $D=11$ supergravity. This procedure will be given explicitly to first order in the coupling constant in these two actions, but it is expected that the procedure generalizes to all orders in the coupling constant as well as to other types of actions constructed from pure spinor superfields \footnote{A similar procedure was used by Chang, Lin, Wang and Yin in \cite{Chang:2014nwa} to find the on-shell solution to abelian and non-abelian $D=10$ supersymmetric Born-Infeld. We thank Martin Cederwall for informing us of their work after the first version of our preprint was submitted.}\label{foot1}.

The procedure consists in using BRST cohomology arguments to define a unique decomposition of the on-shell pure spinor superfield $\Psi(x^m, \theta^\alpha, \lambda^\alpha, \bar\lambda_\alpha, r_\alpha)$ into the sum of two terms as
$$\Psi(x^m, \theta^\alpha, \lambda^\alpha, \bar\lambda_\alpha, r_\alpha) =
\tilde\Psi(x^m, \theta^\alpha, \lambda^\alpha) +
\Lambda( x^m, \theta^\alpha, \lambda^\alpha, \bar\lambda_\alpha, r_\alpha)$$
where $\tilde\Psi(x^m, \theta^\alpha, \lambda^\alpha)$ is independent of the non-minimal variables and $\Lambda$ is constructed from the superfields in $\tilde\Psi$ and the non-minimal variables. Since $\tilde\Psi$ will have a fixed ghost number $g$ ($g$=1 for $D=10$ super-Born-Infeld and $g=3$ for $D=11$ supergravity), it can be expanded as $\tilde\Psi = \lambda^\alpha \tilde A_\alpha (x,\theta)$ or $\tilde\Psi = \lambda^\alpha\lambda^\beta\lambda^\gamma \tilde C_{\alpha\beta\gamma} (x,\theta)$, and it will be shown to first order in the coupling constant that $\tilde A_\alpha (x,\theta)$ and $\tilde C_{\alpha\beta\gamma}(x,\theta)$ correctly describe the on-shell spinor gauge superfield of $D=10$ super-Born-Infeld and the on-shell spinor 3-form superfield of $D=11$ supergravity.

We expect it should be possible to generalize this procedure to all orders in the coupling constant and to other types of pure spinor actions, but there is an important issue concerning these pure spinor actions which needs to be further investigated. If the superfields $\Psi$ in these actions are allowed to have poles of arbitrary order in the non-minimal pure spinor variables, the cohomology arguments used to define the on-shell superfields become invalid. This follows from the well-known property of non-minimal pure spinor variables that one can construct a state $\xi (\lambda, \bar\lambda, \theta, r)$ satisfying $Q\xi =1$ if $\xi$ is allowed to have poles of order $\lambda^{-11}$ in $D=10$ or poles of order 
$\lambda^{-23}$ in $D=11$. And if $\xi$ is allowed in the Hilbert space of states, all BRST cohomology becomes trivial since any state $V$ satisfying $QV=0$ can be expressed as $V = Q (\xi V)$.  

So in order for these actions to correctly describe the on-shell superfields, one needs to impose restrictions on the possible pole dependence of the superfields $\Psi$. But since the pole dependence of the product of superfields can be more singular than the pole dependence of individual superfields, it is not obvious how to restrict the pole dependence of the superfields in a manner which is consistent with the non-linear BRST transformations of the action.

In section 2 of this paper, the $D=10$ pure spinor superparticle and the pure spinor actions for $D=10$ super-Maxwell and super-Yang-Mills will be reviewed. And in section 3, these actions will be generalized to abelian $D=10$ supersymmetric Born-Infeld constructed in terms of a non-minimal pure spinor superfield $\Psi$. 
The super-Born-Infeld equations of motion take the simple form
\begin{eqnarray}
Q\Psi + k(\lambda\gamma^{m}\hat{\chi}\Psi)(\lambda\gamma^{n}\hat{\chi}\Psi)\hat{F}_{mn}\Psi &=& 0
\end{eqnarray}
where $k$ is the dimensionful coupling constant and $\hat\chi^\alpha$ and $\hat F_{mn}$ are operators depending in a complicated manner on the non-minimal variables.
After expanding $\Psi$ in powers of $k$ as $\Psi= \sum_{i=0}^{\infty}k^{i}\Psi_{i}$,
one finds that $\Psi_0$ satisfies the equation $Q\Psi_0=0$ with the super-Maxwell solution $\Psi_0 = \lambda^\alpha A_\alpha (x,\theta)$, and
$\Psi_1$ can be uniquely decomposed as 
$$\Psi_1(x, \theta, \lambda, \bar\lambda, r) =
\tilde\Psi_1(x, \theta, \lambda) +
\Lambda(A_{\alpha}, \lambda, \bar\lambda, r)$$
where $\tilde\Psi_1$ satisfies  \cite{Cederwall:2001td,Chang:2014nwa}
\begin{eqnarray}\label{eq01}
Q\tilde\Psi_1 + (\lambda\gamma^{m}\chi)(\lambda\gamma^{n}\chi){F}_{mn} = 0
\end{eqnarray}
and $\chi^\alpha$ and $F_{mn}$ are the linearized spinor and vector field-strengths constructed from the super-Maxwell superfield in $\Psi_0$. It is straightforward to show that \eqref{eq01} correctly describes the first-order correction of Born-Infeld to the super-Maxwell equations.

In section 4 of this paper, the $D=11$ pure spinor superparticle and the pure spinor action for linearized $D=11$ supergravity will be reviewed. And in section 5, this action will be generalized to the complete $D=11$ supergravity action constructed in terms of a non-minimal pure spinor superfield $\Psi$. 
The supergravity equations of motion take the form
\begin{eqnarray}
Q\Psi + \frac{\kappa}{2}(\lambda\Gamma_{ab}\lambda)R^{a}\Psi R^{b}\Psi + \frac{\kappa}{2}\Psi \{Q , T\}\Psi - \kappa^{2}(\lambda\Gamma_{ab}\lambda)T\Psi R^{a}\Psi R^{b}\Psi &=& 0
\end{eqnarray}
where $\kappa$ is the dimensionful coupling constant and $R^a$ and $T$ are operators depending in a complicated manner on the non-minimal variables.
After expanding $\Psi$ in powers of $\kappa$ as $\Psi= \sum_{i=0}^{\infty}\kappa^{i}\Psi_{i}$,
one finds that $\Psi_0$ satisfies the equation $Q\Psi_0=0$ with the linearized supergravity solution $\Psi_0 = \lambda^\alpha\lambda^\beta\lambda^\gamma C_{\alpha\beta\gamma}(x, \theta)$, and
$\Psi_1$ can be uniquely decomposed as 
$$\Psi_1(x, \theta, \lambda, \bar\lambda, r) =
\tilde\Psi_1(x, \theta, \lambda) +
\Lambda(C_{\alpha\beta\delta}, \lambda, \bar\lambda, r)$$
where $\tilde\Psi_1$ satisfies
\begin{eqnarray}\label{eq02}
Q\tilde\Psi_1 + \frac{1}{2}(\lambda\Gamma_{ab}\lambda)\Phi^{a}\Phi^{b} &=& 0
\end{eqnarray}
and $\Phi^a \equiv \lambda^\alpha E_\alpha^{(0)P}\hat{E}_P{}^a (x,\theta)$ is constructed from the linear deformation of the supergravity supervielbein and the background value of its respective inverse. It is straightforward to show that \eqref{eq02} correctly describes the first-order correction to the linearized supergravity equations.

Finally, Appendices \ref{apA} and \ref{apB} will contain some useful gamma matrix identities in $D=10$ and $D=11$, and Appendix \ref{apC} will explain the relation of the $D=11$ supergravity superfields $\Psi$ and $\Phi^a$.


\vspace{2mm}
\section{Ten-dimensional Pure Spinor Superparticle and Super Yang-Mills}
In this section we will review the pure spinor description for the ten-dimensional superparticle and its connection with ten-dimensional super-Maxwell. We will then discuss the generalization to the non-abelian case. 
\subsection{$D=10$ Pure spinor superparticle}
The ten-dimensional pure spinor superparticle action is given by \cite{Berkovits:2001rb,Berkovits:2002zk}
\begin{eqnarray}
S &=& \int d\tau \left[P_{m}\partial_{\tau}X^{m} + p_{\mu}\partial_{\tau}\theta^{\mu} + w_{\mu}\partial_{\tau}\lambda^{\mu}\right]
\end{eqnarray}
where $X^{m}$ is a ten-dimensional coordinate, $\theta^{\mu}$ is a ten-dimensional Majorana-Weyl spinor, $\lambda^{\mu}$ is a bosonic ten-dimensional Weyl spinor satisfying $\lambda\gamma^{m}\lambda = 0$; and $P_{m}$, $p_{\mu}$, $w_{\mu}$ are the conjugate momenta relative to $X^{m}$, $\theta^{\mu}$, $\lambda^{\mu}$ respectively. We are using Greek/Latin letters from the middle of the alphabet to denote ten-dimensional Majorana-Weyl spinor/vector indices. Furthermore, $(\gamma^{m})^{\mu\nu}$ and $(\gamma^{m})_{\mu\nu}$ are $16\times 16$ symmetric real matrices satisfying $(\gamma^{m})^{\mu\nu}(\gamma^{n})_{\nu\sigma} + (\gamma^{n})^{\mu\nu}(\gamma^{m})_{\nu\sigma} = 2\eta^{mn}\delta^{\mu}_{\sigma}$. The BRST operator is given by
\begin{eqnarray}
Q_{0} &=& \lambda^{\mu}d_{\mu}
\end{eqnarray}
where $d_{\mu} = p_\mu - (\gamma^m \theta)_\mu P_m$ are the fermionic constraints of the $D=10$ Brink-Schwarz superparticle \cite{BRINK1981310}. The physical spectrum is defined as the cohomology of the BRST operator $Q_{0}$. One can show that the ten-dimensional super-Maxwell physical fields are described by ghost number one states: $\Psi = \lambda^{\mu}A_{\mu}$. This can be easily seen since states in the cohomology satisfy the equation of motion and gauge invariance
\begin{eqnarray}
(\gamma^{mnpqr})^{\mu\nu}D_{\mu}A_{\nu} &=& 0\nonumber\\
\delta A_{\mu} &=& D_{\mu}\Lambda\label{eeeq104}
\end{eqnarray}
where $D_\mu = {\partial\over{\partial \theta^\mu}}- (\gamma^m\theta)_\mu \partial_m$.
These are indeed the superspace constraints describing ten-dimensional super-Maxwell. It can be shown that the remaining non-trivial cohomology is found at ghost number 0, 2 and 3 states; describing the super-Maxwell ghost, antifields and antighost, respectively, as dictated by BV quantization.

\subsection{$D=10$ Super-Maxwell}
In order to describe $D=10$ super-Maxwell \eqref{eeeq104} from a well-defined pure spinor action principle, one should introduce non-minimal pure spinor variables \cite{Berkovits:2005bt}. These non-minimal variables were studied in detail in \cite{Berkovits:2006vi} and consist of a pure spinor $\bar{\lambda}_{\mu}$ satisfying $\bar{\lambda}\gamma^{m}\bar{\lambda} = 0$, a fermionic spinor $r_{\mu}$ satisfying $\bar{\lambda}\gamma^{m}r = 0$ and their respective conjugate momenta $\bar{w}^{\mu}$, $s^{\mu}$. The non-minimal BRST operator is defined as $Q = Q_{0} + r_{\mu}\bar{w}^{\mu}$, so that these non-minimal variables will not affect the BRST cohomology. This means that one can always find a representative in the cohomology which is independent of non-minimal variables. 

Note that it will be assumed that the dependence on the non-minimal variables of the states is restricted to diverge slower than $(\lambda\bar\lambda)^{-11}$ when $\lambda^\mu \to 0$. Without this restriction, any BRST-closed operator is BRST-trivial since $Q(\xi V) = V$ where $\xi \equiv (\lambda\bar\lambda +r\theta)^{-1} (\bar\lambda\theta)$. Since the gauge transformation $\delta \Psi = Q\Lambda$  of super-Maxwell is linear, this restriction is easy to enforce by imposing a similar restriction on the gauge parameter $\Lambda$. However, for the non-linear gauge transformations discussed in the following sections for the super-Yang-Mills, supersymmetric Born-Infeld, and supergravity actions, it is unclear how to enforce this restriction. We shall ignore this subtlety here, but it is an important open problem to define the allowed set of states and gauge transformations for $\Psi$ and $\Lambda$ in these nonlinear actions.

\vspace{2mm}
Let $\mathcal{S}_{SM}$ be the following pure spinor action
\begin{eqnarray}\label{eeeq102}
\mathcal{S}_{SM} &=& \int [dZ] \,\Psi Q \Psi
\end{eqnarray}
where $[dZ] = [d^{10}x][d^{16}\theta][d\lambda][d\bar{\lambda}][d r] N$ is the integration measure, $\Psi$ is a pure spinor superfield (which can also depend on non-minimal variables) and $Q$ is the non-minimal BRST-operator. Let us explain what $[dZ]$ means. Firstly, $[d^{10}x][d^{16}\theta]$ is the usual measure on ordinary ten-dimensional superspace. The factors $[d\lambda][d\bar{\lambda}][d r]$ are given by
\begin{eqnarray}
\left[d\lambda\right]\lambda^{\mu}\lambda^{\nu}\lambda^{\rho} &=& (\epsilon T^{-1})^{\mu\nu\rho}_{\hspace{5mm}\sigma_{1}\ldots\sigma_{11}}d\lambda^{\sigma_{1}}\ldots d\lambda^{\sigma_{11}}\nonumber\\
\left[d\bar{\lambda}\right]\bar{\lambda}_{\mu}\bar{\lambda}_{\nu}\bar{\lambda}_{\rho} &=& (\epsilon T)_{\mu\nu\rho}^{\hspace{5mm}\sigma_{1}\ldots\sigma_{11}}d\bar{\lambda}_{\sigma_{1}}\ldots d\bar{\lambda}_{\sigma_{11}}\nonumber\\
\left[d r\right] &=& (\epsilon T^{-1})^{\mu\nu\rho}_{\hspace{5mm}\sigma_{1}\ldots\sigma_{11}}\bar{\lambda}_{\mu}\bar{\lambda}_{\nu}\bar{\lambda}_{\rho}(\frac{\partial}{\partial r_{\sigma_{1}}})\ldots (\frac{\partial}{\partial r_{\sigma_{11}}})
\end{eqnarray}
where the Lorentz-invariant tensors $(\epsilon T)_{\mu\nu\rho}{}^{\sigma_{1}\ldots\sigma_{11}}$ and $(\epsilon T^{-1})^{\mu\nu\rho}{}_{\sigma_{1}\ldots \sigma_{11}}$ were defined in \cite{Berkovits:2006vi}. They are symmetric and gamma-traceless in $(\mu,\nu,\rho)$ and are antisymmetric in $[\sigma_{1},\ldots,\sigma_{11}]$. $N = e^{-Q(\bar{\lambda} \theta)} = e^{(-\bar{\lambda}\lambda - r \theta)}$ is a regularization factor. Since the measure converges as $\lambda^{8}\bar{\lambda}^{11}$ when $\lambda \to 0$, the action is well-defined as long as the integrand diverges slower than $\lambda^{-8}\bar{\lambda}^{-11}$.

\vspace{2mm}
One can easily see that the equation of motion following from \eqref{eeeq102} is given by
\begin{eqnarray}
Q\Psi = 0
\end{eqnarray}
and since the measure factor $[dZ]$ picks out the top cohomology of the ten-dimensional pure spinor BRST operator, the transformation $\delta \Psi = Q\Lambda$ is a symmetry of the action \eqref{eeeq102}. Therefore, \eqref{eeeq102} describes $D=10$ super-Maxwell.

\subsection{$D=10$ Super Yang-Mills}
Let us define $\mathcal{S}_{SYM}$ to be
\begin{eqnarray}\label{eq222}
\mathcal{S}_{SYM} &=& \int [dZ] \,Tr(\frac{1}{2}\Psi Q\Psi + \frac{g}{3}\Psi \Psi \Psi)
\end{eqnarray}
where $[dZ]$ is the measure discussed above, $\Psi$ is a Lie-algebra valued generic pure spinor superfield, $Q$ is the non-minimal BRST operator and $g$ is the coupling constant. For $SU(n)$ gauge group, expand $\Psi$ in the form: $\Psi = \Psi^{a}T^{a}$, where $T^{a}$ are the Lie algebra generators and $a = 1,\ldots , n^{2}-1$. Using the conventions: $[T^{a}, T^{b}] = f^{abc}T^{c}$ with $f^{abc}$ totally antisymmetric, and $Tr(T^{a}T^{b})=\delta^{ab}$, one can rewrite \eqref{eq222} as follows
\begin{eqnarray}
\mathcal{S}_{SYM} &=& \int [dZ](\frac{1}{2}\Psi^{a}Q\Psi^{a} + \frac{g}{6}f^{abc}\Psi^{a}\Psi^{b}\Psi^{c})
\end{eqnarray}
The e.o.m following from this action is given by:
\begin{eqnarray}\label{eq225}
Q\Psi^{a} + \frac{g}{2}f^{abc}\Psi^{b}\Psi^{c} &=& 0
\end{eqnarray}
or in compact form
\begin{eqnarray}\label{eq223}
Q\Psi + g\Psi\Psi &=& 0
\end{eqnarray}
It turns out that \eqref{eq225} is invariant under the BRST symmetry
\begin{eqnarray}
\delta\Psi^{a} &=& Q\Lambda^{a} + f^{abc}\Psi^{b}\Lambda^{c}
\end{eqnarray}
or in compact form
\begin{eqnarray}\label{eq224}
\delta\Psi = Q\Lambda + [\Psi , \Lambda]
\end{eqnarray}
Since the equations \eqref{eq223}, \eqref{eq224} describe on-shell $D=10$ Super Yang-Mills on ordinary superspace \cite{Berkovits:2002zk}, one concludes that the action \eqref{eq222} describes $D=10$ Super Yang-Mills on a pure spinor superspace.

\section{Pure Spinor Description of Abelian Supersymmetric Born-Infeld}
In this section, we review the construction of the pure spinor action for supersymmetric abelian Born-Infeld and deduce the equations of motion on minimal pure spinor superspace to first order in the coupling.
\subsection{Physical operators}
In order to deform the quadratic super-Maxwell action to the supersymmetric Born-Infeld action, Cederwall introduced the ghost number -1 pure spinor operators \cite{Cederwall:2011vy} 
\begin{eqnarray}
\hat{A}_{\mu} &=& -\frac{1}{(\lambda\bar{\lambda})}[\frac{1}{8}(\gamma^{mn}\bar{\lambda})_{\mu}N_{mn} + \frac{1}{4}\bar{\lambda}_{\mu}N]\nonumber\\
\hat{A}_{m} &=& -\frac{1}{4(\lambda\bar{\lambda})}(\bar{\lambda}\gamma_{m}D) + \frac{1}{32(\lambda\bar{\lambda})^{2}}(\bar{\lambda}\gamma_{m}^{\hspace{2mm}np}r)N_{np}\nonumber\\
\hat{\chi}^{\mu} &=& \frac{1}{2(\lambda\bar{\lambda})}(\gamma^{m}\bar{\lambda})^{\mu}\Delta_{m}\nonumber\\
\hat{F}_{mn} &=& -\frac{1}{4(\lambda\bar{\lambda})}(r\gamma_{mn}\hat{\chi}) = \frac{1}{8(\lambda\bar{\lambda})}(\bar{\lambda}\gamma_{mn}^{\hspace{4mm}p}r)\Delta_{p}\label{eeq6}
\end{eqnarray}
where $\Delta_{m}$ is defined by
\begin{eqnarray}
\Delta_{m} &=& \partial_{m} + \frac{1}{4(\lambda\bar{\lambda})}(r\gamma_{m}D) - \frac{1}{32(\lambda\bar{\lambda})^{2}}(r\gamma_{mnp}r)N^{np}\label{eeq5}
\end{eqnarray}

These operators are constructed to satisfy 
\begin{eqnarray}
\left[ Q, \hat{A}_{\mu}\right] &=& -D_{\mu} - 2(\gamma^{m}\lambda)_{\mu}\hat{A}_{m}\nonumber\\
\{Q, \hat{A}_{m}\} &=& \partial_{m} - (\lambda\gamma_{m}\hat{\chi})\nonumber\\
\left[ Q, \hat{\chi}^{\mu}\right] &=& -\frac{1}{2}(\gamma^{mn}\lambda)^{\mu}\hat{F}_{mn}\nonumber\\
\{Q, \hat{F}_{mn}\} &=& 2(\lambda\gamma_{[m}\partial_{n]}\hat{\chi})
\end{eqnarray}
which mimic the superspace equations of motion of $D=10$ Super-Maxwell
\begin{eqnarray}
D_{\alpha}\Psi_0 + Q_{0} A_{\mu} + 2(\gamma^{m}\lambda)_{\mu}A_{m} &=& 0\nonumber\\
\partial_{m}\Psi_0 - Q_{0} A_{m} - (\lambda\gamma_{m}\chi) &=& 0\nonumber\\
Q_{0}\chi^{\mu} + \frac{1}{2}(\lambda\gamma_{mn})^{\mu}F^{mn} &=& 0\nonumber\\
Q_{0}F_{mn} - 2(\lambda\gamma_{[m}\partial_{n]}\chi) &=& 0 \label{eeq1}
\end{eqnarray}
with $\Psi_0 = \lambda^{\mu}A_{\mu}$.

If one acts with these operators on $\Psi_0$, they satisfy
 \begin{eqnarray}
\hat{A}_{\mu}\Psi_0 = A_{\mu}\hspace{2mm},\hspace{2mm}
\hat{A}_{m}\Psi_0 = A_{m}\hspace{2mm},\hspace{2mm}
\hat{\chi}^{\mu}\Psi_0 = \chi^{\mu}\hspace{2mm},\hspace{2mm}
\hat{F}_{mn}\Psi_0 = F_{mn}\label{eeq2}
 \end{eqnarray}
up to BRST-exact terms and certain ``shift-symmetry terms''  defined in \cite{Cederwall:2011vy,Cederwall:2013vba}. 
For example, the operator $\hat{A}_{\mu}$ acts as
\begin{eqnarray}\label{eeq3}
\hat{A}_{\mu}\Psi_0 &=& A_{\mu} - \frac{1}{2(\lambda\bar{\lambda})}(\lambda\gamma^{m})_{\mu}(\bar{\lambda}\gamma_{m}A)
\end{eqnarray}
where the shift symmetry is $\delta A_\mu = (\lambda\gamma_m)_\mu \phi^m$ for any $\phi^m$.
For $\hat{A}_{m}$ one finds that
\begin{eqnarray}
\hat{A}_{m}\Psi_0 &=& A_{m} - (\lambda\gamma_{m}\rho) + Q[\frac{1}{4(\lambda\bar{\lambda})}(\bar{\lambda}\gamma_{m}A)]
\end{eqnarray}
where the on-shell relation $D_{(\mu}A_{\nu)} = -(\gamma^{m})_{\mu\nu}A_{m}$ has been used, $\delta A_m = (\lambda \gamma_m)_\mu \rho^\mu$ is the shift symmetry, and 
\begin{equation}\label{eeq4}
\rho^{\mu} = \frac{1}{2(\lambda\bar{\lambda})}(\bar{\lambda}\gamma^{m})^{\mu}A_{m} + \frac{1}{8(\lambda\bar{\lambda})^{2}}(\bar{\lambda}\gamma^{m})^{\mu}(r\gamma_{m}A)
\end{equation}
Analogously, one can show a similar behavior for the other operators $\hat{\chi}^{\mu}$, $\hat{F}_{mn}$. 


\subsection{$D=10$ Abelian supersymmetric Born-Infeld}
The deformation to the linearized action \eqref{eeeq102} consistent with BRST symmetry is given by \cite{Cederwall:2011vy}
\begin{eqnarray}\label{eeq12}
\mathcal{S}_{SBI} &=& \int [dZ] \left[\frac{1}{2}\Psi Q\Psi + \frac{k}{4}\Psi(\lambda\gamma^{m}\hat{\chi}\Psi)(\lambda\gamma^{n}\hat{\chi}\Psi)\hat{F}_{mn}\Psi\right]
\end{eqnarray}
which is invariant under the BRST transformation
\begin{eqnarray}
\delta\Psi &=& Q\Lambda + k(\lambda\gamma^{m}\hat{\chi}\Psi)(\lambda\gamma^{n}\hat{\chi}\Psi)\hat{F}_{mn}\Lambda + 2k(\lambda\gamma^{m}\hat{\chi}\Psi)(\lambda\gamma^{n}\hat{\chi}\Lambda)\hat{F}_{mn}\Psi
\end{eqnarray}
for any ghost number 0 pure spinor superfield $\Lambda$.  Note that $k$ is a dimensionful parameter related to the string tension by $k = \alpha\ensuremath{'}^{2}$. The equation of motion coming from \eqref{eeq12} is
\begin{eqnarray}\label{eeq8}
Q\Psi + k(\lambda\gamma^{m}\hat{\chi}\Psi)(\lambda\gamma^{n}\hat{\chi}\Psi)\hat{F}_{mn}\Psi &=& 0
\end{eqnarray}
\vspace{2mm}
which can be written in terms of $\Delta_{m}$ as follows
\begin{eqnarray} \label{eeq9}
Q\Psi + \frac{k}{8(\lambda\bar{\lambda})^{2}}(\bar{\lambda}\gamma^{mnp}r)(\Delta_{m}\Psi)(\Delta_{n}\Psi)(\Delta_{p}\Psi) &=& 0\label{eeq9}
\end{eqnarray}

Since the equation of motion of \eqref{eeq9} for $\Psi$ depends explicitly on the non-minimal variables, it is not obvious how to extract from $\Psi$ the Born-Infeld superfield $\tilde A_\mu(x,\theta)$ which should be independent of the non-minimal variables. However, it will now be argued that there is a unique decomposition of the solution to  \eqref{eeq9} as
\begin{eqnarray}\label{eq1000}
\Psi(x,\theta,\lambda,\bar{\lambda},r) &=& \lambda^\mu \tilde A_\mu (x,\theta) + \Lambda (\tilde A_\mu, \lambda, \bar\lambda, r)
\end{eqnarray}
where $\tilde A_\mu(x,\theta)$ is the on-shell Born-Infeld superfield and $\Lambda$ depends on $\tilde A_\mu$ and on the non-minimal variables. This will be explicitly shown here to the leading Born-Infeld correction to super-Maxwell, and work is in progress on extending this to the complete Born-Infeld solution. As mentioned in footnote \ref{foot1}, a similar procedure was used in \cite{Chang:2014nwa} for the abelian and non-abelian Born-Infeld solutions.

To extract this leading-order correction to super-Maxwell from \eqref{eeq9}, we will first expand the pure spinor superfield $\Psi$ in positive powers of $k$:
\begin{eqnarray}\label{eq100}
\Psi(x,\theta,\lambda,\bar{\lambda},r) &=& \sum_{i=0}^{\infty}k^{i}\Psi_{i}
\end{eqnarray}
The replacement of \eqref{eq100} in \eqref{eeq9} gives us the following recursive relations
\begin{eqnarray}
Q\Psi_{0} &=& 0 \label{eq101}\\
Q\Psi_{1} &=& -\frac{1}{8(\lambda\bar{\lambda})^{2}}(\bar{\lambda}\gamma^{mnp}r)\Delta_{m}\Psi_{0}\Delta_{n}\Psi_{0}\Delta_{p}\Psi_{0}\label{eq102}\\
Q\Psi_{2} &=& -\frac{3}{8(\lambda\bar{\lambda})^{2}}(\bar{\lambda}\gamma^{mnp}r)\Delta_{m}\Psi_{1}\Delta_{n}\Psi_{0}\Delta_{p}\Psi_{0}\\
\vdots\nonumber
\end{eqnarray}

To determine $\Lambda$ in \eqref{eq1000}, first note that \eqref{eq101} has the solution $\Psi_0 = \lambda^\mu A_{0\,\mu}$ where $ A_{0\,\mu}$ is the super-Maxwell superfield which is independent of the non-minimal variables.
However, the solution $\Psi_1$ to \eqref{eq102} must depend on the non-minimal variables because the right-hand side of  \eqref{eq102} depends on these variables. To decompose the solution $\Psi_1$ to the form 
\begin{eqnarray}\label{eq1001}
\Psi_1(x,\theta,\lambda,\bar{\lambda},r) &=& \lambda^\mu A_{1\,\mu} (x,\theta) + \Lambda,
\end{eqnarray}
note that 
\eqref{eq102} implies
\begin{eqnarray}\label{eeeq99}
Q(\frac{1}{(\lambda\bar{\lambda})^{2}}(\bar{\lambda}\gamma^{mnp}r)\Delta_{m}\Psi_{0}\Delta_{n}\Psi_{0}\Delta_{p}\Psi_{0}) = 0
\end{eqnarray}
Since any BRST-closed expression can be expressed in terms of minimal variables up to a BRST-trivial term, there must exist a term $\Lambda$ such that 
\begin{eqnarray}\label{eqeqe34}
-\frac{1}{8(\lambda\bar{\lambda})^{2}}(\bar{\lambda}\gamma^{mnp}r)\Delta_{m}\Psi_{0}\Delta_{n}\Psi_{0}\Delta_{p}\Psi_{0} &=& Q\Lambda + F(\Psi_{0})
\end{eqnarray}
where $F({\Psi_{0}})$ is independent of non-minimal variables. This equation determines $\Lambda$ and $F(\Psi_0)$ up to the shift
\begin{eqnarray}\label{eqeqe33}
\delta \Lambda = H(\Psi_0) + Q\Omega, \quad \delta F(\Psi_0) = - Q H(\Psi_0)
\end{eqnarray}
 where $H(\Psi_0)$ only depends on the minimal variables.  But the BRST-trivial shift $F(\Psi_0) \to F(\Psi_0)  - QH(\Psi_0)$ can be cancelled by a redefinition of the field $\Psi_0 \to \Psi_0 - k H(\Psi_0)$.  So the ambiguity in defining $\Lambda$ in  \eqref{eqeqe34} does not affect the physical spectrum.

In order to find $\Lambda$ and $F(\Psi_{0})$ in \eqref{eqeqe34}, first write $\Delta_{m}$ in the more convenient form
\begin{eqnarray}
\Delta_{m} &=& \partial_{m} - \{Q , \hat{A}_{m}\} + \bar{\lambda}\gamma_{m}\hat{\xi}
\end{eqnarray}
where $\hat{\xi}_{\mu}$ is an operator depending on $N_{mn}, D_{\mu}$, etc. Although it is not complicated to determine $\hat{\xi}_{\mu}$, this will not be relevant for our purposes as we will see later. Using the on-shell relation $\partial_{m}A_{\mu} - D_{\mu}A_{m} = (\gamma_{m}\chi)_{\mu}$, one finds that 
\begin{eqnarray}
\Delta_{m}\Psi &=& (\lambda\gamma_{m}\chi) + \lambda\gamma_{m}Q\rho + \bar{\lambda}\gamma_{m}\hat{\xi}\Psi
\end{eqnarray}
So
\begin{eqnarray}
-\frac{1}{8}(\bar{\lambda}\gamma^{mnp}r)\Delta_{m}\Psi_{0}\Delta_{n}\Psi_{0}\Delta_{p}\Psi_{0}&=& -\frac{1}{8(\lambda\bar{\lambda})^{2}}(\bar{\lambda}\gamma^{mnp}r)[(\lambda\gamma_{m}\chi)(\lambda\gamma_{n}\chi)(\lambda\gamma_{p}\chi)\nonumber\\
&& + 3(\lambda\gamma_{m}Q\rho)(\lambda\gamma_{n}\chi)(\lambda\gamma_{p}\chi)\nonumber\\
&& + 3(\lambda\gamma_{m}Q\rho)(\lambda\gamma_{n}Q\rho)(\lambda\gamma_{p}\chi)\nonumber\\
&&+ (\lambda\gamma_{m}Q\rho)(\lambda\gamma_{n}Q\rho)(\lambda\gamma_{p}Q\rho)]
\end{eqnarray}
The first term $H_{1} = -\frac{1}{8(\lambda\bar{\lambda})^{2}}(\bar{\lambda}\gamma_{mnp}r)(\lambda\gamma^{m}\chi)(\lambda\gamma^{n}\chi)(\lambda\gamma^{p}\chi)$ will provide us the term independent of non-minimal variables:
\begin{eqnarray}
H_{1} &=& -\frac{1}{8(\lambda\bar{\lambda})^{2}}(\bar{\lambda}\gamma_{mnp}r)(\lambda\gamma^{m}\chi)(\lambda\gamma^{n}\chi)(\lambda\gamma^{p}\chi)\nonumber\\
&=& \frac{1}{4(\lambda\bar{\lambda})}(r\gamma^{mn}\chi)(\lambda\gamma_{m}\chi)(\lambda\gamma_{n}\chi) - \frac{1}{8(\lambda\bar{\lambda})^{2}}(\lambda r)(\bar{\lambda}\gamma_{mn}\chi)(\lambda\gamma^{m}\chi)(\lambda\gamma^{n}\chi)\nonumber\\
&=& Q[\frac{1}{4(\lambda\bar{\lambda})}(\bar{\lambda}\gamma_{mn}\chi)(\lambda\gamma^{m}\chi)(\lambda\gamma^{n}\chi)] - F_{mn}(\lambda\gamma^{m}\chi)(\lambda\gamma^{n}\chi)
\end{eqnarray}
where the identity \eqref{ap1} was used. Analogous computations show us that the other terms are Q-exact:
\begin{eqnarray}
H_{2} &=& -\frac{3}{8(\lambda\bar{\lambda})^{2}}(\bar{\lambda}\gamma^{mnp}r)(\lambda\gamma_{m}Q\rho)(\lambda\gamma_{n}\chi)(\lambda\gamma_{p}\chi)\nonumber\\
&=& Q[\frac{6}{8(\lambda\bar{\lambda})}(\bar{\lambda}\gamma_{mn}Q\rho)(\lambda\gamma^{m}\chi)(\lambda\gamma^{n}\chi)]
\end{eqnarray}
\begin{eqnarray}
H_{3} &=& -\frac{3}{8(\lambda\bar{\lambda})^{2}}(\bar{\lambda}\gamma^{mnp}r)(\lambda\gamma_{m}Q\rho)(\lambda\gamma_{n}Q\rho)(\lambda\gamma_{p}\chi)\nonumber\\
&=& Q[\frac{6}{8(\lambda\bar{\lambda})}(\bar{\lambda}\gamma_{mn}Q\rho)(\lambda\gamma^{m}Q\rho)(\lambda\gamma^{n}\chi)]
\end{eqnarray}
\begin{eqnarray}
H_{4} &=& -\frac{1}{8(\lambda\bar{\lambda})^{2}}(\bar{\lambda}\gamma^{mnp}r)(\lambda\gamma_{m}Q\rho)(\lambda\gamma_{n}Q\rho)(\lambda\gamma_{p}Q\rho)\nonumber\\
&=& Q[\frac{2}{8(\lambda\bar{\lambda})}(\bar{\lambda}\gamma_{mn}Q\rho)(\lambda\gamma^{m}Q\rho)(\lambda\gamma^{n}Q\rho)]
\end{eqnarray}
Hence, one obtains
\begin{eqnarray}
-\frac{1}{8}(\bar{\lambda}\gamma^{mnp}r)\Delta_{m}\Psi_{0}\Delta_{n}\Psi_{0}\Delta_{p}\Psi_{0} &=& Q[\Lambda] - F_{mn}(\lambda\gamma^{m}\chi)(\lambda\gamma^{n}\chi)
\end{eqnarray}
where $\Lambda$ is defined by the expression
\begin{eqnarray}
\Lambda &=& \frac{1}{4(\lambda\bar{\lambda})}(\bar{\lambda}\gamma_{mn}\chi)(\lambda\gamma^{m}\chi)(\lambda\gamma^{n}\chi) + \frac{3}{4(\lambda\bar{\lambda})}(\bar{\lambda}\gamma_{mn}Q\rho)(\lambda\gamma^{m}\chi)(\lambda\gamma^{n}\chi)\nonumber\\
&& + \frac{3}{4(\lambda\bar{\lambda})}(\bar{\lambda}\gamma_{mn}Q\rho)(\lambda\gamma^{m}Q\rho)(\lambda\gamma^{n}\chi) + \frac{1}{4(\lambda\bar{\lambda})}(\bar{\lambda}\gamma_{mn}Q\rho)(\lambda\gamma^{m}Q\rho)(\lambda\gamma^{n}Q\rho)
\end{eqnarray}
Now, let us define the field $\tilde\Psi = \Psi_{0} + k(\Psi_{1} - \Lambda)$ 
which satisfies to first order in $k$ the equation of motion
\begin{eqnarray}
Q\tilde{\Psi} &=&Q (\Psi_0 + k(\Psi_1 - \Lambda)) =  - k F_{mn}(\lambda\gamma^{m}\chi)(\lambda\gamma^{n}\chi)\label{eeq11}
\end{eqnarray}
where $F_{mn}$, $\chi^{\mu}$ are the usual super-Maxwell superfields constructed from $A_{0\,\mu}$. Since the equation \eqref{eeq11} does not involve non-minimal variables, the solution is
\begin{eqnarray}
\tilde \Psi = \lambda^\mu \tilde {A}_{\mu} 
\end{eqnarray}
where $\tilde A_\mu\equiv A_{0\,\mu} + k A_{1\,\mu}$ satisfies
\begin{eqnarray}
\lambda^{\mu}\lambda^{\nu}\left [D_{\mu}\tilde{A}_{\nu} + k(\gamma^{m}\chi)_{\mu}(\gamma^{n}\chi)_{\nu}F_{mn}\right] &=& 0
\end{eqnarray}
This equation of motion coincides, at first order in $k$, with the abelian supersymmetric Born-Infeld equations of motion \cite{BERGSHOEFF1987371,JAMESGATES1987172,Berkovits:2002ag}. So it has been shown to first order in $k$ that 
\begin{eqnarray}
\Psi &=& \lambda^\mu \tilde A_\mu + k\Lambda
\end{eqnarray}
where $\tilde A_\mu(x, \theta)$ is the on-shell Born-Infeld superfield and $\Lambda$ depends on $A_{0\,\mu}$ and on the non-minimal variables. 

\section{Eleven-Dimensional Pure Spinor Superparticle and Supergravity}
In this section we review the eleven-dimensional pure spinor superparticle and its connection with linearized eleven-dimensional supergravity. 
\subsection{$D=11$ Pure spinor superparticle}
The eleven-dimensional pure spinor superparticle action is given by \cite{Berkovits:2002uc,PhysRevD.97.066002}
\begin{eqnarray}
S &=& \int d\tau \left[P_{m}\partial_{\tau}X^{m} + P_{\mu}\partial_{\tau}\theta^{\mu} + w_{\alpha}(\partial_{\tau}\lambda^{\alpha}+ \partial_\tau Z^M
\Omega_{M\beta}{}^\alpha \lambda^\beta)\right]
\end{eqnarray}
where $X^{m}$ is an eleven-dimensional coordinate, $\theta^{\mu}$ is an eleven-dimensional Majorana spinor, $Z^M=(X^m, \theta^\mu)$, $\lambda^{\alpha}$ is a bosonic eleven-dimensional Majorana spinor satisfying $\lambda\Gamma^{a}\lambda = 0$; $P_{m}$, $P_{\mu}$, $w_{\alpha}$ are the conjugate momenta relative to $X^{m}$, $\theta^{\mu}$, $\lambda^{\alpha}$ respectively, and $\Omega_{M\beta}{}^\alpha$ is the spin connection of the background. We are using Greek/Latin letters from the beginning of the alphabet to denote tangent-space eleven-dimensional spinor/vector indices, and
Greek/Latin letters from the middle of the alphabet to denote coordinate-space eleven-dimensional spinor/vector indices. Furthermore, capital letters from the
beginning of the alphabet will denote tangent-space indices (both spinor and vector) and capital letters from the
middle of the alphabet will denote coordinate-space indices (both spinor and vector). Finally, $(\Gamma^{a})^{\alpha\beta}$ and $(\Gamma^{a})_{\beta\delta}$ are $32\times 32$ symmetric matrices satisfying $(\Gamma^{a})^{\alpha\beta}(\Gamma^{b})_{\beta\delta}$ + $(\Gamma^{b})^{\alpha\beta}(\Gamma^{a})_{\beta\delta} = 2\eta^{ab}\delta^{\alpha}_{\delta}$. The BRST operator is given by
\begin{eqnarray}
Q_{0} &=& \lambda^{\alpha}d_{\alpha}
\end{eqnarray}
where
\begin{eqnarray}
d_\alpha &=& E_{\alpha}^{\hspace{2mm}M} (P_M +\Omega_{M\beta}{}^\gamma w_\gamma \lambda^\beta)
\end{eqnarray}
In a flat Minkowski background, $d_{\alpha} = P_{\alpha} - (\Gamma^{m}\theta)_{\alpha}P_{m}$ are the fermionic constraints of the $D=11$ Brink-Schwarz-like superparticle.

The physical spectrum is defined as the cohomology of the BRST operator $Q_{0}$. One can show that the eleven-dimensional linearized supergravity physical fields are described by ghost number three states: $\Psi = \lambda^{\alpha}\lambda^{\beta}\lambda^{\delta}C_{\alpha\beta\delta}$ \cite{Berkovits:2002uc}
where the physical state condition imposes the following equations of motion and gauge transformations for $C_{\alpha\beta\delta}$
\begin{eqnarray}\label{eqeqe1}
D_{(\alpha}C_{\beta\delta\epsilon)} &=& (\Gamma^{a})_{(\alpha\beta}C_{\vert a\vert \delta\epsilon)}\nonumber\\
\delta C_{\alpha\beta\delta} &=& D_{(\alpha}\Lambda_{\beta\delta)}
\end{eqnarray}
for some superfield $\Lambda_{\beta\delta}$. These are the superspace constraints describing eleven-dimensional linearized supergravity \cite{BRINK1980384}. It can be shown that the remaining non-trivial cohomology is found at ghost number 0, 1, 2, 4, 5, 6 and 7 states; describing the ghosts, antifields and antighosts as dictated by BV quantization of $D=11$ linearized supergravity.

\subsection{$D=11$ Linearized Supergravity}
In order to describe $D=11$ linearized supergravity \eqref{eqeqe1} from a pure spinor action principle, one should introduce eleven-dimensional non-minimal pure spinor variables \cite{Cederwall:2013vba}. These non-minimal variables were studied in detail in \cite{Cederwall:2012es, Cederwall:2009ez} and consist of a pure spinor $\bar{\lambda}_{\alpha}$ satisfying $\bar{\lambda}\Gamma^{a}\bar{\lambda} = 0$, a fermionic spinor $r_{\alpha}$ satisfying $\bar{\lambda}\Gamma^{a}r = 0$ and their respective conjugate momenta $\bar{w}^{\alpha}$, $s^{\alpha}$. The non-minimal BRST operator is defined as $Q = Q_{0} + r_{\alpha}\bar{w}^{\alpha}$, so that these non-minimal variables will not affect the BRST cohomology.

\vspace{2mm}
Let $\mathcal{S}_{LSG}$ be the following pure spinor action
\begin{eqnarray}\label{eeeeq102}
\mathcal{S}_{LSG} &=& \int [dZ] \,\Psi Q \Psi
\end{eqnarray}
where $[dZ] = [d^{11}x][d^{32}\theta][d\lambda][d\bar{\lambda}][d r] N$ is the integration measure, $\Psi$ is a pure spinor superfield (which, in general, can also depend on non-minimal variables) and $Q$ is the non-minimal BRST-operator. Let us explain what $[dZ]$ means. Firstly, $[d^{11}x][d^{32}\theta]$ is the usual measure on ordinary eleven-dimensional superspace. The factors $[d\lambda][d\bar{\lambda}][d r]$ are given by
\begin{eqnarray}
\left[d\lambda\right]\lambda^{\alpha_{1}}\ldots\lambda^{\alpha_{7}} &=& (\epsilon T^{-1})^{\alpha_{1}\ldots\alpha_{7}}_{\hspace{9mm}\beta_{1}\ldots\beta_{23}}d\lambda^{\beta_{1}}\ldots d\lambda^{\beta_{23}}\nonumber\\
\left[d\bar{\lambda}\right]\bar{\lambda}_{\alpha_{1}}\ldots\bar{\lambda}_{\alpha_{7}} &=& (\epsilon T)_{\alpha_{1}\ldots\alpha_{7}}^{\hspace{9mm}\beta_{1}\ldots\beta_{23}}d\bar{\lambda}_{\beta_{1}}\ldots d\bar{\lambda}_{\beta_{23}}\nonumber\\
\left[d r\right] &=& (\epsilon T^{-1})^{\alpha_{1}\ldots\alpha_{7}}_{\hspace{9mm}\beta_{1}\ldots\beta_{23}}\bar{\lambda}_{\alpha_{1}}\ldots\bar{\lambda}_{\alpha_{7}}(\frac{\partial}{\partial r_{\beta_{1}}})\ldots (\frac{\partial}{\partial r_{\beta_{23}}})
\end{eqnarray}
The Lorentz-invariant tensors $(\epsilon T)_{\alpha_{1}\ldots\alpha_{7}}^{\hspace{9mm}\beta_{1}\ldots\beta_{23}}$ and $(\epsilon T^{-1})^{\alpha_{1}\ldots \alpha_{7}}_{\hspace{9mm}\beta_{1}\ldots\beta_{23}}$ were defined in \cite{Cederwall:2009ez}. They are symmetric and gamma-traceless in $(\alpha_{1},\ldots,\alpha_{7})$ and are antisymmetric in $[\beta_{1},\ldots ,\beta_{23}]$. $N$ is a regularization factor which is given by $N = e^{-\lambda\bar{\lambda} - r\theta}$. Since the measure converges as $\lambda^{16}\bar{\lambda}^{23}$ when $\lambda\to 0$, the action is well-defined if the integrand diverges slower than  $\lambda^{-16}\bar{\lambda}^{-23}$.

\vspace{2mm}
One can easily see that the equation of motion following from \eqref{eeeeq102} is given by
\begin{eqnarray}
Q\Psi = 0
\end{eqnarray}
and since the measure factor $[dZ]$ picks out the top cohomology of the eleven-dimensional pure spinor BRST operator, the transformation $\delta \Psi = Q\Lambda$ is a symmetry of the action \eqref{eeeeq102}, that is a gauge symmetry of the theory. Therefore, \eqref{eeeeq102} describes $D=11$ linearized supergravity.


\section{Pure Spinor Description of Complete $D=11$ Supergravity}
As discussed in \cite{Cederwall:2013vba,Cederwall:2010tn} , the pure spinor BRST-invariant action for complete $D=11$ supergravity is given by 
\begin{eqnarray}\label{eq105}
\mathcal{S}_{SG} &=& {1\over{\kappa^2}}\int [dZ][\frac{1}{2}\Psi Q\Psi + \frac{1}{6}(\lambda\Gamma_{ab}\lambda)(1 - \frac{3}{2}T\Psi)\Psi R^{a}\Psi R^{b}\Psi]
\end{eqnarray}
which is invariant under the BRST symmetry 
\begin{eqnarray}
\delta\Psi &=& Q\Lambda + (\lambda\Gamma_{ab}\lambda)R^{a}\Psi R^{b}\Lambda + \frac{1}{2}\Psi\{Q,T\}\Lambda - \frac{1}{2}\Lambda\{Q,T\}\Psi - 2(\lambda\Gamma_{ab}\lambda)T\Psi R^{a}\Psi R^{b}\Lambda\nonumber\\
&& - (\lambda\Gamma_{ab}\lambda)(T\Lambda)R^{a}\Psi R^{b}\Psi
\end{eqnarray}
for any ghost number 2 pure spinor superfield $\Lambda$. Here $\kappa$ is the gravitational coupling constant, and $R^{a}$ and $T$ are ghost number -2 and -3 operators respectively, defined by the relations \cite{Cederwall:2009ez,Cederwall:2010tn}
\begin{eqnarray}
R^{a} &=& -8[\frac{1}{\eta}(\bar{\lambda}\Gamma^{ab}\bar{\lambda})\partial_{b} + \frac{1}{\eta^{2}}(\bar{\lambda}\Gamma^{ab}\bar{\lambda})(\bar{\lambda}\Gamma^{cd}r)(\lambda\Gamma_{bcd}D)\nonumber\\
&& - \frac{4}{\eta^{3}}(\bar{\lambda}\Gamma^{ab}\bar{\lambda})(\bar{\lambda}\Gamma^{cd}r)(\bar{\lambda}\Gamma^{ef}r)(\lambda\Gamma_{fb}\lambda)(\lambda\Gamma_{cde}w)\nonumber\\
&& + \frac{4}{\eta^{3}}(\bar{\lambda}\Gamma^{ac}\bar{\lambda})(\bar{\lambda}\Gamma^{de}r)(\bar{\lambda}\Gamma^{bf}r)(\lambda\Gamma_{fb}\lambda)(\lambda\Gamma_{cde}w)]\label{eeq13}\\
T &=& \frac{512}{\eta^{3}}(\bar{\lambda}\Gamma^{ab}\bar{\lambda})(\bar{\lambda}r)(rr)N_{ab},
\end{eqnarray}
and $\eta \equiv (\lambda \Gamma^{ab}\lambda)(\bar\lambda\Gamma_{ab}\bar\lambda)$.
Note that the action is invariant under the shift symmetry $\delta R^{a} = (\lambda\Gamma^{a}{\cal O})$ for any operator ${\cal O}$.

\vspace{2mm}
The equation of motion coming from the action \eqref{eq105} is
\begin{eqnarray}
Q\Psi + \frac{1}{2}\Psi\{Q,T\}\Psi + \frac{1}{2}(\lambda\Gamma_{ab}\lambda)(1 - 2T\Psi)R^{a}\Psi R^{b}\Psi &=& 0
\end{eqnarray}
To compare with the linearized equations, it is convenient to rescale $\Psi \to \kappa \Psi$ so that $\kappa$ drops out of the quadratic term in the action, and the e.o.m. takes the form
\begin{eqnarray}
Q\Psi + \frac{\kappa}{2}(\lambda\Gamma_{ab}\lambda)R^{a}\Psi R^{b}\Psi + \frac{\kappa}{2}\Psi \{Q , T\}\Psi - \kappa^{2}(\lambda\Gamma_{ab}\lambda)T\Psi R^{a}\Psi R^{b}\Psi &=& 0\label{eq106}
\end{eqnarray}  
In order to find the superspace equations of motion, we expand the pure spinor superfield $\Psi$ in positive powers of $\kappa$
\begin{eqnarray}
\Psi &=& \sum_{n=0}^{\infty} \kappa^{n}\Psi_{n}
\end{eqnarray}
where $\Psi_{0}$ is the linearized solution satisfying $Q\Psi_{0} = 0$, which describes linearized 11D supergravity. The recursive relations that one finds from equation \eqref{eq106} are:
\begin{eqnarray}
Q\Psi_{0} &=& 0\\
Q\Psi_{1} + \frac{1}{2}(\lambda\Gamma_{ab}\lambda)R^{a}\Psi_{0} R^{b}\Psi_{0} + \frac{1}{2}\Psi_{0} \{Q , T\}\Psi_{0} &=& 0\label{eq107}\\
\vdots\nonumber
\end{eqnarray}

The procedure will now be the same as that applied to the Born-Infeld case: We will first write the non-minimal contribution to \eqref{eq107} as a BRST-exact term $Q\Lambda$. We will then define a new superfield $\tilde\Psi = \Psi-\Lambda$, which will satisfy the equation $Q\tilde\Psi = G(\Psi_{0})$ where $G(\Psi_{0})$ is independent of non-minimal variables. We will finally identify $\tilde C_{\alpha\beta\gamma} = C_{0\,\alpha\beta\gamma} + \kappa C_{1\,\alpha\beta\gamma}$ in $\tilde \Psi = \lambda^\alpha\lambda^\beta \lambda^\gamma \tilde C_{\alpha\beta\gamma}$ as the first-order correction to the linearized D=11 superfield.

\vspace{2mm}
To find $\Lambda$ and $G(\Psi_0)$, the first step will be to write $R^{a}\Psi_{0}$ in terms of a superfield $\Phi^a(x, \theta, \lambda)$
depending only on minimal variables as 
\begin{eqnarray}\label{eeq15}
R^{a}\Psi_{0} &=& \Phi^{a}(x, \theta, \lambda) + Q(f^{a}) + \lambda\Gamma^a{\cal O}
\end{eqnarray} 
where $\lambda \Gamma^a {\cal O}$ is the shift symmetry of $R^a$. To linearized order in the supergravity deformation of the background, the superfield $\Phi^a$ can be expressed in terms of the super-vielbein $E_A{}^P$ and
its inverse $E_P{}^A$ as
\begin{eqnarray}\label{eeqq19}
\Phi^a &=& \lambda^\alpha E_\alpha^{(0)P} \hat E_P{}^a
\end{eqnarray}
where $E_A{}^P$ and $E_P{}^A$ have been expanded around their background values $\hat E_A{}^P$ and $\hat E_P{}^A$ as
\begin{eqnarray}
E_A{}^P &=& \hat E_A{}^P + \kappa E_A^{(0)P} + \kappa^2 E_A^{(1)P} + ...,\nonumber\\
E_P{}^{A} &=& \hat E_P{}^A + \kappa E_P^{(0)A} + \kappa^2 E_P^{(1)A} + ... .
\end{eqnarray}
For example, if one is expanding around the Minkowski space background, $\hat E_a{}^p = \delta_a^p$, $\hat E_\alpha{}^\mu = \delta_\alpha^\mu$ and $\hat E_\alpha{}^m = -
(\Gamma^m\theta)_\alpha$. Note that $E^{(0)P}_\alpha \hat E_P{}^a + \hat E_\alpha{}^P E^{(0) a}_P =0$, so one can also express $\Phi^a$ to linearized order in the deformation as 
\begin{eqnarray}\label{eeqq20}
\Phi^a &=& - \lambda^\alpha \hat E_\alpha{}^P E_P^{(0)a}
\end{eqnarray}
 
Since all of the supergravity fields are contained in $\Psi_0$, one should be able to describe $\Phi$ in terms of $\Psi_0$.  As discussed in \cite{Cederwall:2009ez}, this relation is given by \eqref{eeq15} and it will be explicitly shown in Appendix \ref{apC} that
\begin{eqnarray}
f^{a} &=& -\frac{24}{\eta^{2}}(\bar{\lambda}\Gamma^{ab}\bar{\lambda})(\bar{\lambda}\Gamma^{cd}r)(\lambda\Gamma_{bcd})^{\delta}C_{\delta\alpha\beta}\lambda^{\alpha}\lambda^{\beta} - \frac{24}{\eta}(\bar{\lambda}\Gamma^{ab}\bar{\lambda})C_{b\alpha\beta}\lambda^{\alpha}\lambda^{\beta}
\end{eqnarray}

Plugging eq. \eqref{eeq15} in \eqref{eq107} implies that
\begin{eqnarray}
Q\Psi_{1} + \frac{1}{2}(\lambda\Gamma_{ab}\lambda)[\Phi^{a} + Qf^{a}][\Phi^{b} + Qf^{b}] - Q\left[\frac{1}{2}\Psi_{0}T\Psi_{0}\right] &=& 0,
\end{eqnarray}
which implies that 
\begin{eqnarray}\label{eeq19}
Q(\Psi_1 - \Lambda) = - \frac{1}{2}(\lambda\Gamma_{ab}\lambda)\Phi^{a}\Phi^{b}
\end{eqnarray}
where
$$\Lambda =  \frac{1}{2}\Psi_{0}T\Psi_{0} + (\lambda\Gamma_{ab}\lambda)\Phi^{a}f^{b} - \frac{1}{2}(\lambda\Gamma_{ab}\lambda)f^{a}Qf^{b}.$$

Hence one can define the superfield $\tilde{\Psi}$:
\begin{eqnarray}
\tilde{\Psi} = \Psi_{0} + \kappa(\Psi_1 - \Lambda)
\end{eqnarray}
which will satisfy the following e.o.m at linear order in $\kappa$
\begin{eqnarray}
Q\tilde{\Psi} &=& - \frac{\kappa}{2}(\lambda\Gamma_{ab}\lambda)\Phi^{a}\Phi^{b}
\end{eqnarray}
which implies
\begin{eqnarray}\label{eq99}
\lambda^{\alpha}\lambda^{\beta}\lambda^{\delta}\lambda^{\epsilon}
[D_{\alpha}\tilde{C}_{\beta\delta\epsilon} +\frac{\kappa}{2}(\Gamma_{ab})_{\alpha\beta}E_{\delta}^{(0)P} \hat E_P{}^a E_{\epsilon}^{(0)Q}\hat E_Q{}^b] =0 
\end{eqnarray}
where $\tilde\Psi = \lambda^\alpha\lambda^\beta\lambda^\delta \tilde C_{\alpha\beta\delta}$.

This equation of motion \eqref{eq99} will now be shown to coincide with the $D=11$ supergravity equations of motion at first order in $\kappa$. The non-linear 
$D=11$ supergravity equations of motion can be expressed using pure spinors as 
\begin{eqnarray}\label{eqeqe181}
\lambda^\alpha \lambda^\beta \lambda^\gamma \lambda^\delta H_{\alpha\beta\gamma\delta} =0
\end{eqnarray}
where we
use the standard transformation rule from curved to tangent-space indices for the 4-form superfield strength:
\begin{eqnarray}\label{eqeqe301}
H_{\alpha\beta\delta\epsilon} &=& E_{\alpha}^{\hspace{2mm}M}E_{\beta}^{\hspace{2mm}N}E_{\delta}^{\hspace{2mm}P}E_{\epsilon}^{\hspace{2mm}Q}H_{MNPQ}
\end{eqnarray}
and $H_{MNPQ} = \nabla_{[M} C_{NPQ]}$. Furthermore, \eqref{eqeqe301} implies that one can choose conventional constraints (by appropriately defining $C_{\alpha\beta a}$ and $C_{\alpha a b}$) so that 
 $$H_{\alpha\beta\gamma\delta}= 
 H_{\alpha\beta\gamma a}= 0, \quad
 H_{\alpha\beta a b}= -{1\over{12}} (\Gamma_{ab})_{\alpha\beta}.$$
This is expected since
there are no physical supergravity fields
with the dimensions of $H_{\alpha\beta\gamma\delta}$, $H_{\alpha\beta\gamma a}$ and $H_{\alpha\beta a b}$. 
 

To perform an expansion in $\kappa$ and compare with \eqref{eq99}, define
\begin{eqnarray}
\hat{H}_{\alpha\beta\gamma\delta} &=& \hat{E}_{\alpha}^{\hspace{2mm}M}\hat{E}_{\beta}^{\hspace{2mm}N}\hat{E}_{\gamma}^{\hspace{2mm}P}\hat{E}_{\delta}^{\hspace{2mm}Q}H_{MNPQ}.
\end{eqnarray}\label{eqeq000} 
Equation
\eqref{eqeqe181} implies that
$$
0 = \lambda^\alpha \lambda^\beta \lambda^\gamma \lambda^\delta (\hat H_{\alpha\beta\gamma\delta} +4\kappa \hat E_\alpha{}^M \hat E_\beta{}^N \hat E_\gamma{}^P E_\delta^{(0)Q}
H_{MNPQ} $$
\begin{eqnarray}
+6 
\kappa^2 \hat E_\alpha{}^M \hat E_\beta{}^N E_\gamma^{(0) P} E_\delta{}^{(0)Q} H_{MNPQ} + ...)
 \end{eqnarray}\label{eqq180}
 $$= \lambda^\alpha \lambda^\beta \lambda^\gamma \lambda^\delta (\hat H_{\alpha\beta\gamma\delta} +4\kappa \hat E_\alpha{}^M \hat E_\beta{}^N \hat E_\gamma{}^P E_\delta^{(0)Q}
E_M{}^A E_N{}^B E_P{}^C E_Q{}^D H_{ABCD}$$
\begin{eqnarray}
+6 
\kappa^2 \hat E_\alpha{}^M \hat E_\beta{}^N E_\gamma^{(0) P} E_\delta^{(0)Q} E_M{}^A E_N{}^B E_P{}^C E_Q{}^D H_{ABCD} + ...)
 \end{eqnarray}
 $$= \lambda^\alpha \lambda^\beta \lambda^\gamma \lambda^\delta (\hat H_{\alpha\beta\gamma\delta} +12\kappa^2  \hat E_\gamma{}^P E_P^{(0) a} E_\delta^{(0)Q}\hat E_Q{}^b
 H_{\alpha\beta a b}$$
 \begin{eqnarray}
+6 
\kappa^2  E_\gamma^{(0) P}\hat E_P{}^a E_\delta^{(0)Q} \hat E_Q{}^b H_{\alpha \beta a b} + ...)
 \end{eqnarray}
 \begin{eqnarray}\label{eqq22}
&=& 
 \lambda^\alpha \lambda^\beta \lambda^\gamma \lambda^\delta (\hat H_{\alpha\beta\gamma\delta}  + {1\over 2} 
\kappa^2  E_\gamma^{(0) P} \hat E_P{}^a E_\delta^{(0)Q} \hat E_Q{}^b (\Gamma_{ab})_{\alpha \beta} + ...)
\end{eqnarray}
 where $...$ denotes terms higher-order in $\kappa$.  Since 
 $$\lambda^\alpha \lambda^\beta \lambda^\gamma \lambda^\delta\hat H_{\alpha\beta\gamma\delta} = \kappa \lambda^\alpha \lambda^\beta \lambda^\gamma \lambda^\delta D_\alpha \tilde C_{\beta\gamma\delta},$$
 equation \eqref{eqq22} for the
 back-reaction to $\hat H_{\alpha\beta\gamma\delta}$ coincides with \eqref{eq99}.

\section{Acknowledgments}
MG acknowledges FAPESP grant 2015/23732-2 for financial support and NB acknowledges FAPESP grants 2016/01343-7 and 2014/18634-9 and CNPq grant 300256/94-9 for partial financial support.

\begin{appendices}
\section{$D=10$ gamma matrix identities}\label{apA}
In $D=10$ dimensions, one has chiral and antichiral spinors which have been denoted here by $\chi^{\mu}$ and $\chi_{\mu}$ respectively. The product of two spinors can be decomposed into two forms depending on the chiralities of the spinors used:
\begin{eqnarray}
\xi_{\mu}\chi^{\nu} &=& \frac{1}{16}\delta^{\nu}_{\mu}(\xi\chi) - \frac{1}{2!16}(\gamma^{mn})^{\nu}_{\hspace{2mm}\mu}(\xi\gamma_{mn}\chi) + \frac{1}{4!16}(\gamma^{mnpq})^{\nu}_{\hspace{2mm}\mu}(\xi\gamma_{mnpq}\chi)\\
\xi^{\mu}\chi^{\nu} &=& \frac{1}{16}\gamma_{m}^{\mu\nu}(\xi\gamma^{m}\chi) + \frac{1}{3!16}(\gamma_{mnp})^{\mu\nu}(\xi\gamma^{mnp}\chi) + \frac{1}{5!32}\gamma^{\mu\nu}_{mnpqr}(\xi\gamma^{mnpqr}\chi) \label{ap4}
\end{eqnarray}
The 1-form and 5-form are symmetric, and the 3-form is antisymmetric. Furthermore, it is true that $(\gamma^{mn})^{\mu}_{\hspace{2mm}\nu} = -(\gamma^{mn})_{\nu}^{\hspace{2mm}\mu}$, $(\gamma^{mnpq})^{\mu}_{\hspace{2mm}\nu} = (\gamma^{mnpq})_{\nu}^{\hspace{2mm}\mu}$.

\vspace{2mm}
Two particularly useful identities are:
\begin{eqnarray}
(\gamma^{m})_{(\mu\nu}(\gamma_{m})_{\rho)\sigma}  &=& 0\\
(\gamma^{m})^{\mu}_{\hspace{2mm}\nu}(\gamma_{m})^{\rho}_{\hspace{2mm}\sigma} &=& 4(\gamma^{m})^{\mu\rho}(\gamma_{m})_{\nu\sigma} - 2\delta^{\mu}_{\nu}\delta^{\rho}_{\sigma} - 8\delta^{\mu}_{\sigma}\delta^{\rho}_{\nu} \label{ap1}
\end{eqnarray}
From \ref{ap1} we can deduce the following:
\begin{eqnarray}
(\gamma^{mn})^{\mu}_{\hspace{2mm}\nu}\gamma_{mnp}^{\rho\sigma} &=& 2(\gamma^{m})^{\mu\rho}(\gamma_{pm})^{\sigma}_{\hspace{2mm}\nu} + 6\gamma_{p}^{\mu\rho}\delta^{\sigma}_{\nu} - (\rho \leftrightarrow \sigma)\label{ap2} \\
(\gamma^{mn})^{\mu}_{\hspace{2mm}\nu}(\gamma_{mnp})_{\rho\sigma} &=& -2\gamma^{m}_{\nu\sigma}(\gamma_{pm})^{\mu}_{\hspace{2mm}\rho} + 6(\gamma_{p})_{\nu\sigma}\delta^{\mu}_{\rho} - (\rho \leftrightarrow \sigma)\\
\gamma_{mnp}^{\mu\nu}(\gamma^{mnp})^{\rho\sigma} &=& 12[\gamma_{m}^{\mu\sigma}(\gamma^{m})^{\nu\rho} - \gamma_{m}^{\mu\rho}(\gamma^{m})^{\nu\sigma}] \label{ap5}\\
\gamma^{\mu\nu}_{mnp}\gamma^{mnp}_{\rho\sigma} &=& 48(\delta^{\mu}_{\rho}\delta^{\nu}_{\sigma} - \delta^{\mu}_{\sigma}\delta^{\nu}_{\rho})\label{ap6}
\end{eqnarray}

\section{$D=11$ gamma matrix identities}\label{apB}
In $D=11$ dimensions, one has Majorana spinors and an antisymmetric tensor $C_{\alpha\beta}$ (and its inverse) which can be used to raise and lower spinor indices. The product of two spinors can be decomposed into the form
\begin{eqnarray}\label{app7}
\chi^{\alpha}\psi^{\beta} &=& -\frac{1}{32}C^{\alpha\beta}(\chi\psi) + \frac{1}{32}(\Gamma^{a})^{\alpha\beta}(\chi\Gamma_{a}\psi) - \frac{1}{2!.32}(\Gamma^{ab})^{\alpha\beta}(\chi\Gamma_{ab}\psi) + \frac{1}{3!.32}(\Gamma^{abc})^{\alpha\beta}(\chi\Gamma_{abc}\psi)\nonumber\\
&& - \frac{1}{4!.32}(\Gamma^{abcd})^{\alpha\beta}(\chi\Gamma_{abcd}\psi) + \frac{1}{5!.32}(\Gamma^{abcde})^{\alpha\beta}(\chi\Gamma_{abcde}\psi)
\end{eqnarray}
The 1-form, 2-form and 5-form are symmetric; and the 0-form, 3-form and 4-form are antisymmetric.

\vspace{2mm}
The crucial identity in eleven dimensions is
\begin{eqnarray}\label{ap7}
(\Gamma^{ab})_{(\alpha\beta}(\Gamma_{b})_{\delta\epsilon)} &=& 0
\end{eqnarray}

\vspace{2mm}
One can find analogous formulae to \eqref{ap1}-\eqref{ap6} for $D=11$ dimensions. However, they do not enter into any computations of this paper, therefore we will not list them.

\vspace{2mm}
From \eqref{ap7} and the pure spinor constraint, one can find the following useful pure spinor identities
\begin{eqnarray}
(\bar{\lambda}\Gamma^{ab}\bar{\lambda})(\Gamma_{b}\bar{\lambda})_{\alpha} &=& 0 \label{app1}\\
(\bar{\lambda}\Gamma^{[ab}\bar{\lambda})(\bar{\lambda}\Gamma^{c]d}\bar{\lambda}) &=& 0 \label{app2}\\
(\bar{\lambda}\Gamma^{[ab}\bar{\lambda})(\bar{\lambda}\Gamma^{cd]}\bar{\lambda}) &=& 0 \label{app3}\\ 
(\bar{\lambda}\Gamma^{[ab}\bar{\lambda})(\bar{\lambda}\Gamma^{cd]}r) &=& 0 \label{app4}
\end{eqnarray}

If $a$ is a shift-symmetry index, there exists a very useful identity which states the following
\begin{eqnarray}\label{ap8}
(\bar{\lambda}\Gamma^{ab}\bar{\lambda})(\lambda\Gamma_{cb}\lambda) &=& \frac{1}{2}\delta^{a}_{c}\eta
\end{eqnarray}
This can be easily seen from the following argument. Eqn. \eqref{ap7} implies the relation
\begin{eqnarray*}
-(\bar{\lambda}\Gamma^{ab}\bar{\lambda})(\lambda\Gamma_{b}\Gamma_{c}\lambda) &=& 2(\bar{\lambda}\Gamma^{ab}\bar{\lambda})(\bar{\lambda}\Gamma_{b}\Gamma_{c}\lambda) + 2(\bar{\lambda}\Gamma^{ab}\Gamma_{c}\lambda)(\bar{\lambda}\Gamma_{c}\lambda)
\end{eqnarray*}
which can be rewritten in the more convenient form
\begin{eqnarray*}
-(\bar{\lambda}\Gamma^{ab}\bar{\lambda})(\lambda\Gamma_{b}\Gamma_{c}\lambda) &=& \lambda\Gamma^a\xi_c + 4\delta^a_c(\lambda\bar{\lambda})^2 - 4\delta^a_c (\bar{\lambda}\Gamma^b \lambda)(\bar{\lambda}\Gamma_b \lambda)
\end{eqnarray*}
where $\xi_c^\alpha$ is defined as follows
\begin{eqnarray*}
\xi_c^\alpha &=& -2(\bar{\lambda}\Gamma_c)^\alpha (\bar\lambda \lambda) - 2(\bar{\lambda}\Gamma^b)^\alpha (\bar{\lambda}\Gamma_b \Gamma_c \lambda) + 2(\bar{\lambda}\Gamma_{bc})^\alpha (\lambda\Gamma^b \bar{\lambda}) + 4\bar{\lambda}^\alpha (\lambda\Gamma_c \bar\lambda)
\end{eqnarray*}
The use of \eqref{app7} allows us to write
\begin{eqnarray*}
-(\lambda\bar{\lambda})^2 &=& -\frac{1}{64}\eta + \frac{1}{3840}(\lambda\Gamma^{abcde}\lambda)(\bar{\lambda}\Gamma_{abcde}\bar{\lambda})\\
(\bar{\lambda}\Gamma^a \lambda)(\bar{\lambda}\Gamma_a \lambda) &=& -\frac{7}{64}\eta - \frac{1}{3840}(\lambda\Gamma^{abcde}\lambda)(\bar{\lambda}\Gamma_{abcde}\bar{\lambda})
\end{eqnarray*}
Therefore,
\begin{eqnarray*}
(\bar{\lambda}\Gamma^{ab}\bar{\lambda})(\lambda\Gamma_{cb}\lambda) &=& \frac{1}{2}\delta^a_c \eta + \lambda\Gamma^a\xi_c
\end{eqnarray*}

\section{Relation between $\Psi$ and $\Phi^{a}$}\label{apC}
At linearized level, there exists a simple relation between $\Psi$ and $\Phi^{a}$. To find this relation, define
\begin{eqnarray}\label{eqeqe18}
\hat {H}_{ABCD} &=& \hat{E}_{A}^{\hspace{2mm}M}\hat{E}_{B}^{\hspace{2mm}N}\hat{E}_{C}^{\hspace{2mm}P}\hat{E}_{D}^{\hspace{2mm}Q} {H}_{MNPQ}
\end{eqnarray}
as in \eqref{eqeq000}. Using the conventions
\begin{eqnarray}\label{eeq17}
H_{\alpha\beta\delta\gamma} = 0\hspace{2mm},\hspace{2mm}H_{a\alpha\beta\delta} = 0\hspace{2mm},\hspace{2mm}H_{ab\alpha\beta} = -\frac{1}{12}(\Gamma_{ab})_{\alpha\beta}\hspace{2mm},\hspace{2mm}H_{abc\alpha} = 0,
\end{eqnarray}
one finds that
\begin{eqnarray}
\lambda^\alpha \lambda^\beta\lambda^\gamma\hat{H}_{a\alpha\beta\gamma} &=& \lambda^\alpha \lambda^\beta\lambda^\gamma\hat E_a{}^M \hat E_\alpha{}^N \hat E_\beta{}^P \hat E_\gamma{}^Q E_M{}^A E_N{}^B E_P{}^C E_Q{}^D H_{ABCD}
\end{eqnarray}
\begin{eqnarray}
&=& 3\kappa \lambda^\alpha \lambda^\beta\lambda^\gamma \hat E_\alpha{}^N E_N^{(0)b} 
H_{ab \beta\gamma} + ...
\end{eqnarray}
\begin{eqnarray}
&=&  \frac{1}{4}\kappa \Phi^b \lambda^\beta\lambda^\gamma  
(\Gamma_{ab})_{\beta\gamma} + ...
\end{eqnarray}
where $...$ denotes terms of order $\kappa^2$ and $\Phi^b = -\lambda^\alpha \hat E_\alpha{}^N E_N^{(0)b}$.

Since 
$$\lambda^\alpha\lambda^\beta\lambda^\gamma\hat H_{a\alpha\beta\gamma} =
\kappa(\partial_a \Psi_0 - 3Q(\lambda^{\alpha}\lambda^\beta C_{a\alpha\beta})) ,$$
one obtains the relation
\begin{eqnarray}\label{eqeqe305}
\partial_{a}\Psi_{0} &=& \frac{1}{4}(\lambda\Gamma_{ab}\lambda)\Phi^{b} + 3Q(\lambda^{\alpha}\lambda^\beta C_{a\alpha\beta})
\end{eqnarray}

\vspace{2mm}
The use of equation \eqref{eqeqe305} and the linearized e.o.m
\begin{eqnarray}\label{eeq20}
D_{\alpha}\Psi_{0} + 3Q_{0}(C_{\alpha\beta\delta})\lambda^{\beta}\lambda^{\delta} = -6(\Gamma^{a}\lambda)_{\alpha}C_{a\beta\delta}\lambda^{\beta}\lambda^{\delta}
\end{eqnarray}
allows us to compute the action of $R^{a}$ on $\Psi_{0}$ in the form displayed in \eqref{eeq15}. To see this, it will be useful to express $R^{a}$ in the more convenient way \cite{Cederwall:2010tn}
\begin{eqnarray}
R^{a} &=& -8[\frac{1}{\eta}(\bar{\lambda}\Gamma^{ab}\bar{\lambda})\partial_{b} + \frac{1}{\eta^{2}}(\bar{\lambda}\Gamma^{ab}\bar{\lambda})(\bar{\lambda}\Gamma^{cd}r)(\lambda\Gamma_{bcd}D)\nonumber\\
&& - \{Q , \frac{1}{\eta^{2}}(\bar{\lambda}\Gamma^{ab}\bar{\lambda})(\bar{\lambda}\Gamma^{cd}r)\}(\lambda\Gamma_{bcd}w)]
\end{eqnarray} 
Therefore,
\begin{eqnarray}
R^{a}\Psi_{0} &=& -8[\frac{1}{\eta}(\bar{\lambda}\Gamma^{ab}\bar{\lambda})\partial_{b}\Psi_{0} + \frac{1}{\eta^{2}}(\bar{\lambda}\Gamma^{ab}\bar{\lambda})(\bar{\lambda}\Gamma^{cd}r)(\lambda\Gamma_{bcd}D\Psi_{0})\nonumber\\
&& + 3\{Q , \frac{1}{\eta^{2}}(\bar{\lambda}\Gamma^{ab}\bar{\lambda})(\bar{\lambda}\Gamma^{cd}r)\}(\lambda\Gamma_{bcd})^{\alpha}C_{\alpha\beta\delta}\lambda^{\beta}\lambda^{\delta}]\nonumber\\
&=& -8[\frac{1}{\eta}(\bar{\lambda}\Gamma^{ab}\bar{\lambda})\partial_{b}\Psi_{0} - \frac{3}{\eta^{2}}(\bar{\lambda}\Gamma^{ab}\bar{\lambda})(\bar{\lambda}\Gamma^{cd}r)(\lambda\Gamma_{bcd})^{\alpha} (Q C_{\alpha\beta\delta})\lambda^{\beta}\lambda^{\delta}\nonumber\\
&& - \frac{6}{\eta^{2}}(\bar{\lambda}\Gamma^{ab}\bar{\lambda})(\bar{\lambda}\Gamma^{cd}r)(\lambda\Gamma_{bcd}\Gamma^{e}\lambda)C_{e\alpha\beta}\lambda^{\alpha}\lambda^{\beta}\nonumber\\
&& + 3\{Q , \frac{1}{\eta^{2}}(\bar{\lambda}\Gamma^{ab}\bar{\lambda})(\bar{\lambda}\Gamma^{cd}r)\}(\lambda\Gamma_{bcd})^{\alpha}C_{\alpha\beta\delta}\lambda^{\beta}\lambda^{\delta}]\nonumber\\
&=& -8\{\frac{1}{\eta}(\bar{\lambda}\Gamma^{ab}\bar{\lambda})\partial_{b}\Psi_{0} + Q\left[ \frac{3}{\eta^{2}}(\bar{\lambda}\Gamma^{ab}\bar{\lambda})(\bar{\lambda}\Gamma^{cd}r)(\lambda\Gamma_{bcd})^{\alpha}C_{\alpha\beta\delta}\lambda^{\beta}\lambda^{\delta}\right]\nonumber\\
&& - \frac{6}{\eta^{2}}(\bar{\lambda}\Gamma^{ab}\bar{\lambda})(\bar{\lambda}\Gamma^{cd}r)[-2(\lambda\Gamma_{bd}\lambda)\eta^{e}_{c} + (\lambda\Gamma_{cd}\lambda)\eta^{e}_{b}]C_{e\alpha\beta}\lambda^{\alpha}\lambda^{\beta}\}\nonumber\\
&=& -8\{\frac{1}{\eta}(\bar{\lambda}\Gamma^{ab}\bar{\lambda})\partial_{b}\Psi_{0} + Q\left[ \frac{3}{\eta^{2}}(\bar{\lambda}\Gamma^{ab}\bar{\lambda})(\bar{\lambda}\Gamma^{cd}r)(\lambda\Gamma_{bcd})^{\alpha}C_{\alpha\beta\delta}\lambda^{\beta}\lambda^{\delta}\right]\nonumber\\
&& + \frac{6}{\eta}(\bar{\lambda}\Gamma^{ab}r)C_{b\alpha\beta}\lambda^{\alpha}\lambda^{\beta} - \frac{6}{\eta^{2}}(\bar{\lambda}\Gamma^{ab}\bar{\lambda})
 (\bar{\lambda}\Gamma^{cd}r)(\lambda\Gamma_{cd}\lambda) C_{b\alpha\beta}\lambda^{\alpha}\lambda^{\beta}\}\nonumber\\
&=& -8\{\frac{1}{\eta}(\bar{\lambda}\Gamma^{ab}\bar{\lambda})\partial_{b}\Psi_{0} + Q\left[ \frac{3}{\eta^{2}}(\bar{\lambda}\Gamma^{ab}\bar{\lambda})(\bar{\lambda}\Gamma^{cd}r)(\lambda\Gamma_{bcd})^{\alpha}C_{\alpha\beta\delta}\lambda^{\beta}\lambda^{\delta}\right]\nonumber\\
&& + Q[\frac{3}{\eta}(\bar{\lambda}\Gamma^{ab}\bar{\lambda})C_{b\alpha\beta}\lambda^{\alpha}\lambda^{\beta}] - \frac{3}{\eta}(\bar{\lambda}\Gamma^{ab}\bar{\lambda})Q[C_{b\alpha\beta}\lambda^{\alpha}\lambda^{\beta}]\}\nonumber\\
&=& -\frac{2}{\eta}(\bar{\lambda}\Gamma^{ab}\bar{\lambda})(\lambda\Gamma_{bc}\lambda)\Phi^{c} + Q\left[ -\frac{24}{\eta^{2}}(\bar{\lambda}\Gamma^{ab}\bar{\lambda})(\bar{\lambda}\Gamma^{cd}r)(\lambda\Gamma_{bcd})^{\alpha}C_{\alpha\beta\delta}\lambda^{\beta}\lambda^{\delta} - \frac{24}{\eta}(\bar{\lambda}\Gamma^{ab}\bar{\lambda})C_{b\alpha\beta}\lambda^{\alpha}\lambda^{\beta}\right]\nonumber\\
&=& \Phi^{a} + Q\left[ -\frac{24}{\eta^{2}}(\bar{\lambda}\Gamma^{ab}\bar{\lambda})(\bar{\lambda}\Gamma^{cd}r)(\lambda\Gamma_{bcd})^{\alpha}C_{\alpha\beta\delta}\lambda^{\beta}\lambda^{\delta} - \frac{24}{\eta}(\bar{\lambda}\Gamma^{ab}\bar{\lambda})C_{b\alpha\beta}\lambda^{\alpha}\lambda^{\beta}\right]
\end{eqnarray}
Notice that in order for the normalization factor of $\Phi^{a}$ to be one after applying $R^{a}$ on $\Psi$, one should choose the conventions used for $R^{a}$ in \eqref{eeq13} and those displayed in \eqref{eeq17}.
\end{appendices}

\providecommand{\href}[2]{#2}\begingroup\raggedright\endgroup


\begin{thebibliography}{10}

\bibitem{HOWE1991141}
P.~Howe, ``Pure spinor lines in superspace and ten-dimensional supersymmetric
  theories,''
  \href{http://dx.doi.org/https://doi.org/10.1016/0370-2693(91)91221-G}{{\em
  Physics Letters B} {\bf 258} (1991) no.~1, 141 -- 144}.
  \url{http://www.sciencedirect.com/science/article/pii/037026939191221G}.

\bibitem{Berkovits:2001rb}
N.~Berkovits, ``{Covariant quantization of the superparticle using pure
  spinors},'' \href{http://dx.doi.org/10.1088/1126-6708/2001/09/016}{{\em JHEP}
  {\bf 09} (2001)  016},
\href{http://arxiv.org/abs/hep-th/0105050}{{\tt arXiv:hep-th/0105050
  [hep-th]}}.

\bibitem{Berkovits:2002uc}
N.~Berkovits, ``{Towards covariant quantization of the supermembrane},''
  \href{http://dx.doi.org/10.1088/1126-6708/2002/09/051}{{\em JHEP} {\bf 09}
  (2002)  051},
\href{http://arxiv.org/abs/hep-th/0201151}{{\tt arXiv:hep-th/0201151
  [hep-th]}}.

\bibitem{Gomez:2013sla}
H.~Gomez and C.~R. Mafra, ``{The closed-string 3-loop amplitude and
  S-duality},'' \href{http://dx.doi.org/10.1007/JHEP10(2013)217}{{\em JHEP}
  {\bf 10} (2013)  217},
\href{http://arxiv.org/abs/1308.6567}{{\tt arXiv:1308.6567 [hep-th]}}.

\bibitem{Cederwall:2010tn}
M.~Cederwall, ``{D=11 supergravity with manifest supersymmetry},''
  \href{http://dx.doi.org/10.1142/S0217732310034407}{{\em Mod. Phys. Lett.}
  {\bf A25} (2010)  3201--3212},
\href{http://arxiv.org/abs/1001.0112}{{\tt arXiv:1001.0112 [hep-th]}}.

\bibitem{Cederwall:2013vba}
M.~Cederwall, ``{Pure spinor superfields -- an overview},''
  \href{http://dx.doi.org/10.1007/978-3-319-03774-5_4}{{\em Springer Proc.
  Phys.} {\bf 153} (2014)  61--93},
\href{http://arxiv.org/abs/1307.1762}{{\tt arXiv:1307.1762 [hep-th]}}.

\bibitem{Cederwall:2011vy}
M.~Cederwall and A.~Karlsson, ``{Pure spinor superfields and Born-Infeld
  theory},'' \href{http://dx.doi.org/10.1007/JHEP11(2011)134}{{\em JHEP} {\bf
  11} (2011)  134},
\href{http://arxiv.org/abs/1109.0809}{{\tt arXiv:1109.0809 [hep-th]}}.

\bibitem{Chang:2014nwa}
C.-M. Chang, Y.-H. Lin, Y.~Wang, and X.~Yin, ``{Deformations with Maximal
  Supersymmetries Part 2: Off-shell Formulation},''
  \href{http://dx.doi.org/10.1007/JHEP04(2016)171}{{\em JHEP} {\bf 04} (2016)
  171},
\href{http://arxiv.org/abs/1403.0709}{{\tt arXiv:1403.0709 [hep-th]}}.

\bibitem{Cederwall:2001td}
M.~Cederwall, B.~E.~W. Nilsson, and D.~Tsimpis, ``{D = 10 superYang-Mills at
  O(alpha-prime**2)},''
  \href{http://dx.doi.org/10.1088/1126-6708/2001/07/042}{{\em JHEP} {\bf 07}
  (2001)  042},
\href{http://arxiv.org/abs/hep-th/0104236}{{\tt arXiv:hep-th/0104236
  [hep-th]}}.

\bibitem{Berkovits:2002zk}
N.~Berkovits, ``{ICTP lectures on covariant quantization of the superstring},''
  {\em ICTP Lect. Notes Ser.} {\bf 13} (2003)  57--107,
\href{http://arxiv.org/abs/hep-th/0209059}{{\tt arXiv:hep-th/0209059
  [hep-th]}}.

\bibitem{BRINK1981310}
L.~Brink and J.~H. Schwarz, ``Quantum superspace,''
  \href{http://dx.doi.org/https://doi.org/10.1016/0370-2693(81)90093-9}{{\em
  Physics Letters B} {\bf 100} (1981) no.~4, 310 -- 312}.
  \url{http://www.sciencedirect.com/science/article/pii/0370269381900939}.

\bibitem{Berkovits:2005bt}
N.~Berkovits, ``{Pure spinor formalism as an N=2 topological string},''
  \href{http://dx.doi.org/10.1088/1126-6708/2005/10/089}{{\em JHEP} {\bf 10}
  (2005)  089},
\href{http://arxiv.org/abs/hep-th/0509120}{{\tt arXiv:hep-th/0509120
  [hep-th]}}.

\bibitem{Berkovits:2006vi}
N.~Berkovits and N.~Nekrasov, ``{Multiloop superstring amplitudes from
  non-minimal pure spinor formalism},''
  \href{http://dx.doi.org/10.1088/1126-6708/2006/12/029}{{\em JHEP} {\bf 12}
  (2006)  029},
\href{http://arxiv.org/abs/hep-th/0609012}{{\tt arXiv:hep-th/0609012
  [hep-th]}}.

\bibitem{BERGSHOEFF1987371}
E.~Bergshoeff, M.~Rakowski, and E.~Sezgin, ``Higher derivative super
  {Yang}-{Mills} theories,''
  \href{http://dx.doi.org/https://doi.org/10.1016/0370-2693(87)91017-3}{{\em
  Physics Letters B} {\bf 185} (1987) no.~3, 371 -- 376}.
  \url{http://www.sciencedirect.com/science/article/pii/0370269387910173}.

\bibitem{JAMESGATES1987172}
S.~J. Gates and S.~Vashakidze, ``On d = 10, n = 1 supersymmetry, superspace
  geometry and superstring effects (i),''
  \href{http://dx.doi.org/https://doi.org/10.1016/0550-3213(87)90470-6}{{\em
  Nuclear Physics B} {\bf 291} (1987)  172 -- 204}.
  \url{http://www.sciencedirect.com/science/article/pii/0550321387904706}.

\bibitem{Berkovits:2002ag}
N.~Berkovits and V.~Pershin, ``{Supersymmetric Born-Infeld from the pure spinor
  formalism of the open superstring},''
  \href{http://dx.doi.org/10.1088/1126-6708/2003/01/023}{{\em JHEP} {\bf 01}
  (2003)  023},
\href{http://arxiv.org/abs/hep-th/0205154}{{\tt arXiv:hep-th/0205154
  [hep-th]}}.

\bibitem{PhysRevD.97.066002}
M.~Guillen, \href{http://dx.doi.org/10.1103/PhysRevD.97.066002}{``Equivalence
  of the 11d pure spinor and {Brink}-{Schwarz}-like superparticle
  cohomologies,''{\em Phys. Rev. D} {\bf 97} (Mar, 2018)  066002}.
  \url{https://link.aps.org/doi/10.1103/PhysRevD.97.066002}.

\bibitem{BRINK1980384}
L.~Brink and P.~Howe, ``Eleven-dimensional supergravity on the mass shell in
  superspace,''
  \href{http://dx.doi.org/https://doi.org/10.1016/0370-2693(80)91002-3}{{\em
  Physics Letters B} {\bf 91} (1980) no.~3, 384 -- 386}.
  \url{http://www.sciencedirect.com/science/article/pii/0370269380910023}.

\bibitem{Cederwall:2012es}
M.~Cederwall and A.~Karlsson, ``{Loop amplitudes in maximal supergravity with
  manifest supersymmetry},''
  \href{http://dx.doi.org/10.1007/JHEP03(2013)114}{{\em JHEP} {\bf 03} (2013)
  114},
\href{http://arxiv.org/abs/1212.5175}{{\tt arXiv:1212.5175 [hep-th]}}.

\bibitem{Cederwall:2009ez}
M.~Cederwall, ``{Towards a manifestly supersymmetric action for 11-dimensional
  supergravity},'' \href{http://dx.doi.org/10.1007/JHEP01(2010)117}{{\em JHEP}
  {\bf 01} (2010)  117},
\href{http://arxiv.org/abs/0912.1814}{{\tt arXiv:0912.1814 [hep-th]}}.

\end{thebibliography}
\end{document}